**Abstract**

The search for the origin of cosmic rays is a quest of almost a hundred years. A recent theoretical proposal gives quantitative predictions, which can be tested with data. Specifically, it has been suggested, that all cosmic rays can be attributed to just three source sites: i) supernova explosions into the interstellar medium, ii) supernova explosions into a stellar wind, and iii) powerful radiogalaxies. The cosmic rays from any extragalactic source suffer from interaction with the microwave background, leading to the Greisen-Zatsepin-Kuzmin cutoff. While the particle energies, the spectrum and the chemical composition of cosmic rays over the energy range from about GeV to about 100 EeV can be interpreted in the theory, there are exciting measurements now: New measurements show that there are cosmic ray events beyond the Greisen-Zatsepin-Kuzmin cutoff. We discuss here possible sources, and specifically ask whether powerful radiogalaxies are suitable candidates.

# The origin of cosmic rays

Peter L. Biermann
Max Planck Institut für Radioastronomie
Auf dem Hügel 69, D-53121 Bonn, Germany

December 10, 1994

## 1 INTRODUCTION

Ever since their discovery in 1912 by Hess and Kohlhörster the origin of cosmic rays has been one of the main enigmas of physics; the energies of the particles observed far exceed the energies even in the most powerful planned accelerators on earth. A classic book is by Hayakawa (1969), and an important recent book is by Berezinskii *et al.* (1990). Discovering a clue about the origin of cosmic rays may help us to explore very high energy physics of particles as well as cosmic accelerators.

Most of the lower energy cosmic rays are believed to be due to the explosion of stars into the normal interstellar medium (Lagage & Cesarsky 1983) and some of the higher energy cosmic rays to explosions of massive stars into their stellar wind (Völk & Biermann, 1988). For the energy range beyond a few $10^{15}$ eV there is general dispute about the origin of the particles.

Various models (see, *e.g.*, Axford 1994) exist for the high energy spectrum of cosmic rays beyond the "knee", where the spectrum shows a slight turn-down:

1. A postulated galactic wind (Jokipii & Morfill 1987) may accelerate particles at a galactic wind termination shock; this is argued to contribute particles over the entire range of particle energies, from low energies to the end of the galactic cosmic ray spectrum.

2. The multiple shocks in the environment of OB superbubbles and young supernova remnants (Bykov & Toptygin 1990, 1992, Polcaro *et al.* 1991, 1993, Bykov & Fleishman 1992, Ip & Axford 1992) may also contribute.

3. The cosmic background of active galactic nuclei may contribute through the production of energetic neutrons which convert back to protons (Protheroe & Szabo 1992). This is a model to account mainly for the sharpness of the spectral turnover near a few $10^{15}$ eV.

4. Many authors have tried time and again to associate the the high energy cosmic rays with pulsars (a relevant short discussion is given by Hillas 1984).

For all such models, a clear prediction of the spectrum and its chemical abundance distribution, as well as a detailed check with the airshower size distribution, both for slanted and oblique showers, is desirable. All these models can possibly explain the data to about a few $10^{18}$ eV, but have difficulty explaining any further extent of the observed cosmic ray spectrum.

## 1.1 A model for the origin of cosmic rays

Recently we have proposed that the origin of cosmic rays can be explained as arising from three sources:

1. Supernova explosions into the interstellar medium,

2. Supernova explosions into a predecessor stellar wind,

3. The hot spots of giant powerful radio galaxies.

The theory makes specific and quantitative predictions for the origin of the energies of the particles, for the spectrum, and for the chemical composition.

This proposal has been developed, tested and described over the particle energy range from about GeV to about 100 EeV (1 EeV = $10^{18}$ eV) in Biermann (1993a), Biermann & Cassinelli (1993), Biermann & Strom (1993), Stanev *et al.* (1993), Rachen & Biermann (1993), Rachen *et al.* (1993), Nath & Biermann (1993, 1994a, b), Biermann *et al.* 1994, in Biermann (1993b, c, 1994, 1995) and various shorter reports.

## 1.2 The extreme energy events

The interaction of high energy particles with the microwave background leads to the theoretically well established Greisen-Zatsepin-Kuzmin (GZK)-cutoff near about 50 EeV. For the recent flurry of data on the highest energies (Efimov *et al.* 1988, 1992, Bird *et al.* 1994b, Hayashida *et al.* 1994), which appear to have energies beyond the GZK-cutoff, various models have been put forward:

- 1. A new version of a galactic halo model (Wdowczyk *et al.* 1994), where the halo accelerates heavy nuclei, that then get scattered in the Galactic magnetic field. As support the authors mention that the Fly's Eye event lies in the Galactic plane; we note that the Yakutsk event is very close to the Fly's Eye event on the sky. As a difficulty the authors see the electron component, since the model requires a fairly strong halo magnetic field, which could lead to excessive synchrotron radiation.

- 2. Powerful radio galaxies (*e.g.* Biermann & Strittmatter 1987, Rachen & Biermann 1993) are candidates, and have the advantage that all candidates ought to be known; here the difficulties are threefold: First, there is no reasonable nucleus which can survive photo-disintegration over the distance from any very powerful radio galaxy; second, in the direction of the events there seems to be no *famous* candidate radiogalaxy (Elbert & Sommers 1994, Hayashida *et al.* 1994); and third, the Fly's Eye event as well as the Yakutsk event lie so close to the Galactic plane, that few reliable extragalactic data exist.

- 3. Cosmological defects (*e.g.* Bhattacharjee 1991, Sigl *et al.* 1994) have the advantage, that little is known about them, so many things are possible. One difficulty with this model is that it requires two cosmic ray source components beyond the energy 3 EeV, where the observed chemical composition and the spectrum change (Bird *et al.* 1994a), because cosmological defects cannot explain all data from 3 EeV to the newest events.

- 4. Obviously, a weakly interacting particle of high energy may come to us from a very distant source.

At the Tokyo Workshop many of these data and the instruments used to obtain them have been discussed in some detail (Nagano 1994, Wdowczyk & Wolfendale 1994, Sommers 1994). We concentrate on the highest energy here, and first discuss the propagation, second the possible source sites, third the extragalactic source candidates, and finally fourth, the extreme energy events.

## 2 A MINIMAL MODEL FOR THE INTERGALACTIC MAGNETIC FIELD

### 2.1 The magnetic field topology

To understand the propagation of very high energy cosmic rays through the universe we require knowledge of the intergalactic magnetic fields (Kronberg 1994): Using rotation measure data from polarized extragalactic radio sources, this gives an upper limit of $10^{-9}$ Gauss for any prevailing field, assumed to have a comoving length scale of about 1 Mpc. Since we do not know the origin of any presumed intergalactic magnetic field, we require recourse to a minimal model:

Galactic evolution naturally leads to a minimum hypothesis to account for the topology of the intergalactic magnetic field: Galactic winds from normal and starburst galaxies carry out low energy cosmic rays as well as magnetic fields. The low energy cosmic rays can reionize the intergalactic medium (Nath & Biermann 1993). This model for reionization suggests that the epoch of reionization was after the first galaxies and quasars formed, and so at a moderately low redshift; this should be testable soon. Along with the low energy cosmic rays magnetic fields are carried out from the ubiquitous normal galaxies, and become very weak upon expansion. Since the expansion speed of any galactic wind is of order the escape speed from galaxies, it is cosmologically slow, and so the spread in distance cannot cover all space easily. As a result the galactic winds produce a topology of intergalactic magnetic fields, which is commensurate with the large scale structure of the galaxy distribution. Therefore, in this model, the spatial arrangement of the intergalactic field is composed of walls, filaments, nodes and voids with the same spatial length scales as the galaxy distribution.

For the high energy cosmic rays this then leads naturally to a mean free path for scattering of order 30 - 100 Mpc, which ensures that very high energy cosmic rays can come to us very nearly on straight paths. As a bonus, they thus minimize their losses against the microwave background. If true, this immediately suggests, that we might to be able to "observe" the sources.

## 2.2 The large scale structure

The bubble structure of the galaxy distribution is visible in the nearby universe as a sheet of galaxies along the supergalactic plane (de Vaucouleurs 1956, 1975a, b, 1976). Nearby radio galaxies also distribute themselves along this plane (Shaver & Pierre 1989, Shaver 1991). Using the 1-Jy catalogue of bright radiosources at 5 GHz (Kühr *et al.* 1981) and the 3CR catalogue (Laing *et al.* 1983, Spinrad *et al.* 1985) and limiting the sample to all radiosources below redshift 0.02, one indeed sees again the strong tendency to cluster along the supergalactic plane; in the sample of strong radio sources we exclude normal galaxies here such as the starburst galaxy M82. The sample can be extended by also including the brightest obscured sources in the galactic plane, which are likely to be extragalactic (to, say, 30 Jy at 178 MHz, this adds another 4 sources), however have no known redshift. We have ignored very weak nearby sources such as 0149+35 and 0120+33, both of which have redshift 0.016 (Heckman *et al.* 1994; Eilek, priv.comm.) The total sample constructed in such a way thus consists of 23 radio sources; many of these sources show well known jets (Bridle & Perley 1984), such as NGC315, 3C31, 3C449 and others; the obscured radio source 3C134 has no known redshift, but shows the classical shape of an edge bright radio galaxy (Macdonald *et al.* 1968, Leahy & Williams 1984).

For high energy cosmic rays all this means, that those particles which can cross the bubble size, may come from the powerful source candidates in all parts of the sky, but as soon as the maximum distance the particles can travel drops below the bubble size, the candidate sources should begin to cluster along the supergalactic plane. This entails in the picture in which the sources are, *e.g.*, powerful radiogalaxies, that the event directions on the sky should begin to cluster along the supergalactic plane near and beyond the GZK-cutoff.

As a consequence radio sources, that are expected to be also nearby sources of cosmic rays, should show this clustering, whether individually known or not. The events from Fly's Eye, Yakutsk and Akeno are all fairly near to the supergalactic plane, consistent with such a hypothesis.

## 2.3 The particle orbits in our Galaxy

As, *e.g.*, shown by Giller *et al.* (1993) and Chi *et al.* (1994), high energy particles that come into our Galaxy from the outside do experience some distortion in their orbits due to the Galactic magnetic field. The real magnitude of this effect, obviously, depends on the rigidity of the particles, or in other words, the energy over charge ratio for highly relativistic particles. This may make it difficult to locate the cosmic ray sources on the sky.

# 3 THE CLASS OF EXTRAGALACTIC SOURCE CANDIDATES

Powerful radiogalaxies produce a spectrum of cosmic rays that is modified by the microwave background and can dominate the cosmic ray spectrum observed on earth above 3 EeV (Rachen & Biermann 1993, Rachen *et al.* 1993). The transition from the heavy nuclei dominated galactic component – due to stellar wind explosions – to a proton dominated extragalactic component leads to a simultaneous change in spectral shape and chemical composition. Although it is often claimed that the distribution of ultra high energy cosmic ray events shows an enhancement in the galactic disk, this is not in contradiction to an extragalactic origin; we note that Cyg A and some other of the strongest extragalactic sources are in the galactic plane direction. However, the usually large distance of powerful radio galaxies restricts the prediction range of the model to energies below the GZK-cutoff, and we have to check to what extent it can explain the observed events above that limit.

Which extragalactic sources should be best as candidates for the production of high energy cosmic ray particles? We have argued previously, that powerful radio galaxies, that show hot spots in their radio emission, such as Cyg A, are suitable candidates (Rachen & Biermann 1993, Rachen *et al.* 1993). The most powerful radio galaxies show the edge bright morphology, usually called Fanaroff Riley class II (Fanaroff & Riley 1974), or short FR II. There is the opposite class of radio galaxies which are center bright (Miley 1980), such as 3C449, called FR I. And then there many radio galaxies which are not easily attributed to either class, such as M87 and NGC315, which are asymmetric and seem to be between the two classes. There is evidence that sources such as M87 also accelerate nuclei to very high energies (Biermann & Strittmatter 1987).

We do not know whether in these unclear cases or maybe even in FR I radio galaxies particles are accelerated sufficiently far outside the central source in such a way as to minimize adiabatic losses on the path to the intergalactic medium. We will assume in the following that particle acceleration in clear FR I radio galaxies, such as 3C449, is strongest near the center, and so that in those cases adiabatic losses diminish their contribution for the feeding of the cosmic ray population in true intergalactic space.

The remaining radio galaxies we will refer to as cosmic ray source candidates, and thus we exclude sources such as Cen A or 3C449, but include sources such as M87 or NGC315, and certainly include sources such as Cyg A, the classical FR II source.

# 4 INDIVIDUAL CANDIDATE SOURCES FOR THE UHE COSMIC RAYS

## 4.1 The highest energy events

The most recent three extremely high energy events, beyond $10^{20}$ eV, from Fly's Eye, Akeno and Yakutsk pose a new riddle, since they are beyond the GZK-cutoff. The Fly's Eye and Akeno events have been described in Bird *et al.* (1994b) and in Hayashida *et al.* (1994). The Yakutsk event is given in Efimov *et al.* (1988, 1992).

The location on the sky for the Fly's Eye event is in right ascension and declination (referred to 1950) $85.2 \pm 0.4$, $48.0^{+5.2}_{-6.3}$ degrees, for the Yakutsk event $75.2 \pm 7.5$, $45.5 \pm 4$, and for the Akeno event $18.9 \pm 1$, $21.1 \pm 1$; here we have taken the error box data from the original publications for the Fly's Eye and Akeno events, and the error box size estimate for the Yakutsk event from Sigl *et al.* (1994).

The interaction with the microwave background gives the maximum distances of suitable candidates, and is slightly dependent on nuclear mass.

The events may be explained by heavy nuclei of originally iron mass (Rachen *et al.*, in prep.). Nuclei heavier than iron even may explain the determined energy, but hardly the high flux sometimes deduced from the events. Iron can hardly be more than a small fraction of the total cosmic ray flux from radio galaxies. In addition, the maximum distance from which iron particles of 300 EeV at earth can originate, is below 3 Mpc, thus even less than the distance to the starburst galaxy M82. Only if the energy of these extreme events is systematically overestimated by a factor of about two, which appears possible within the given errorbars (at the several $\sigma$ level), then candidate sources like NGC315 become viable. In that case a straight path to us is possible. For sources such as M87 with a very low redshift a bent path is required.

Protons have the advantage, that their energy loss is relatively large per interaction, and so the fluctuations can be large. Poisson statistics suggest, that with a small probability protons can originate at a large distance (see Sigl *et al.* 1994). Thus, with a probability reduced by three standard deviations, protons which arrive with 170 EeV may originate at a distance of redshift z =

0.016 to 0.02 ($H_o$ assumed to be 75 km/sec/Mpc); the implied original energy is 400 to 500 EeV. To arrive with 320 EeV from such distances requires an original energy of 1000 to 1200 EeV, which stretches any model for the possible source physics. It follows that an interpretation as protons may require nearby sources such as M87, and so a large bending of the path to us is implied. Obviously, as noted above, if the energy of the events is systematically overestimated, then the range of options increases.

If an angular correlation between cosmic ray events and sources with known redshift on the sky could be established both below and above the nominal GZK-cutoff, this would lead to limits on the intergalactic magnetic field. Assuming that we see Fe nuclei above the GZK-cutoff, and protons below, and can both identify with sources at redshift of order 0.02, we obtain an upper limit for the intergalactic magnetic field of order $10^{-11}$ Gauss for a nearly homogeneous field, and somewhat higher limits for a tangled field. Assuming, on the other hand, that we see protons both below and above the GZK-limit, with the same redshift range again, then the limit is relaxed to $3\ 10^{-10}$ Gauss in the homogeneous case. Assuming, third, that we see an angular correlation below, but not above the GZK-cutoff, then in this model we do require Fe nuclei above the GZK-cutoff, and the magnetic field can be estimated directly to about $5\ 10^{-11}$ Gauss in the homogeneous case, since then the Fe nuclei presumably come from nearby sources such as M87.

It has been claimed variously, that there is no powerful radiogalaxy in the direction of these extreme events. However, the angular distance which we allow here between the direction to a source candidate, and the arrival direction of a very energetic event, is a free parameter which we have to guess, depending on the strength of the intergalactic or galactic halo magnetic fields. With order of 20 degrees we have the following candidates:

For the Fly's Eye event there is the radio source 3C134 (the position is 5h 1m 17.73s, +38d 2m 6.1s, MacDonald *et al.* 1968; Leahy *et al.* 1984, Spinrad *et al.* 1985); the radio structure of this source resembles that of a powerful radio galaxy, but there is no redshift known because no optical counterpart was found yet due to heavy obscuration by galactic dust, and the argument that this is an extragalactic source is based on the radiostructure. Assuming its real size is on the lower bound of the typical range for powerful radiogalaxies, its distance can be as low as 30 Mpc, making it an acceptable candidate.

The Yakutsk event is rather close to the Fly's Eye event, so we suggest the same source for this event.

For the Akeno event there is the radio galaxy NGC 315 (its position is 0h 55m 5.64s +30d 4m 56.8s, z = 0.016, Kühr *et al.* 1981). The difficulty with this radio galaxy is that it is not a paradigm for Fanaroff Riley class II, the edge brightened radio galaxies, that have inspired us up to now; it has the size of the biggest FR-II's and well collimated jets, but no bright hot spots, and thus may or may not be a powerful source for energetic particles in the intergalactic medium. Another radio galaxy in this part of the sky is 3C31 (position 1h 4m 41.64s +32d 8m 7.5s, z=0.017, Kühr *et al.* 1981). 3C31 is a center bright source and sits in a small group of early type galaxies; above we excluded such clean cases of center bright radio sources as acceptable cosmic ray source candidates.

Most, even unknown radio galaxies as cosmic ray source candidates beyond the GZK-cutoff are expected to be quite close to the supergalactic plane: the Akeno event is almost on the supergalactic plane, while the error boxes for the other two events reach to distances less than 20 degrees of the supergalactic plane.

If we disregard all ideas about angular correlation on the sky (*i.e.* assuming strong magnetic fields), then obviously Cen A is the closest radio galaxy, but it is center brightened, and thus not intrinsically as powerful as a more distant radio galaxy such as, *e.g.*, Cyg A with z=0.0565. The asymmetric structure of M87 at z=0.0043 leaves it as a candidate.

We are thus left with still the same choices for an origin of the highest energy cosmic rays, beyond the GZK-cutoff. As regards the option of a weakly interacting particle we note the following coincidences:

The radioquasar 3C147 (position 05h 38m 43.51s, 49d 49m 42.8s, z= 0.545, Spinrad *et al.* 1985) is in the error range of both the Fly's Eye and the Yakutsk event, as already noted by Sigl *et al.* (1994); there is no other suitable candidate known (Hewitt & Burbidge 1993) directly in the error box. In the error box of the Akeno event there is the optically selected quasar PG 0117+213 (position 1d 17m 34.7s, 21d 18m 4s, z = 1.493; Schmidt & Green 1983; Spinrad *et al.* 1985; Hewitt & Burbidge 1987, 1989, 1993), which is weakly seen at radiowavelengths (Kellermann *et al.* 1989), not detected in the farinfrared (Sanders *et al.* 1989), and also not seen in X-rays (Wilkes *et al.* 1994). This quasar is very luminous, with an estimated accretion disk luminosity of $2 \cdot 10^{47}$ erg/sec (Falcke *et al.* 1994).

# 5 SUMMARY

For the cosmic rays we propose three sites of origin: i) Explosions of supernovae into the interstellar medium. This provides for all Hydrogen observed and energetic electrons up to about 30 GeV. ii) Explosions of supernovae into stellar winds. This gives Helium and heavier elements from GeV particle energies as well as electrons above about 30 GeV. This component is heavily enriched beyond the knee, which is a feature in the spectra of cosmic rays accelerated in stellar winds due to a reduced drift energy gain. iii) Near 3 EeV there is a rapid switch in chemical composition to Hydrogen and Helium and the dominant source may become the cosmological population of powerful radiogalaxies.

The new extreme energy events can be explained by heavy nuclei arising from radio galaxies and radio quasars only, if the the energy for the events is systematically overestimated; this is consistent with the error bars given. The large fluctuations in the energy loss of protons make them a viable alternative. Various other models have been also suggested. All models for the events beyond 50 EeV face severe and challenging difficulties to explain the observations.

So far, the model for the origin of the cosmic rays has passed a number of tests against observations, allowing first quantitative checks to be made. Further work will concentrate on the transport of cosmic rays, the secondary to primary ratio, and the farinfrared-radio emission correlation of galaxies. No proposed model to explain the events beyond 50 EeV can be excluded yet.

# 6  ACKNOWLEDGEMENTS

This work is the result of a close collaboration with Drs. J. Rachen and T. Stanev. I wish to thank my collaborators in cosmic ray work Drs. J.P. Cassinelli, H. Falcke, T.K. Gaisser, J.R. Jokipii, H. Meyer, B. Nath, J. Rachen, E.-S. Seo, T. Stanev, R.G. Strom, and B. Wiebel as well as Drs. V.S. Berezinsky, P. Bhattacharjee, A. Bykov, J. cronin, L. Dedenko, V. Dogiel, M. Giller, J. Eilek, P. Evenson, C. Jarlskog, J.F. McKenzie, M. Nagano, M.I. Pravdin, V.S. Ptuskin, V.V. Usov, J. Wdowczyk, and G. Zank for various helpful communications and many intense discussions. High Energy Physics with the author is supported by a NATO travel grant (CRG 9100072).